\documentstyle[prl,aps]{revtex}

\begin{document}
\title{Field Theoretical Approach to Electrochemical Deposition\thanks{%
Supported by CAPES-Brazil}}
\author{H. A. Taroco, A. L. Mota\thanks{%
motaal@ufsj.edu.br}}
\address{{\normalsize Departamento de Ci\^{e}ncias Naturais, Universidade Federal de}%
\\
S\~{a}o Jo\~{a}o del Rei\\
{\normalsize Caixa Postal 110, CEP 36.301-160, S\~ao Jo\~ao del Rei, MG,}\\
Brazil}
\maketitle

\begin{abstract}
In this work we present an application of the $\lambda\phi^4$ field
theoretical model to the adsorption of atoms and molecules on metallic
surfaces - the electrochemical deposition. The usual approach to this system
consists in the computational simulation using Monte Carlo techniques of an
effective lattice-gas Hamiltonian. We construct an effective model towards a
comparison between the lattice-gas Hamiltonian and the discrete version of
the $\lambda\phi^4$ Hamiltonian, obtaining the relationships between the
model parameters and electrochemical quantities. The $\lambda\phi^4$ model
is studied in the mean field approximation, and the results are fitted and
compared to numerical simulated and experimental data.
\end{abstract}

%
%

\section{Introduction}

%
%

There is a recent convergence on the subjects of studies and techniques of
electrochemistry and surface science. In particular, the processes where
ions and molecules are deposited over a metallic surface can be treated with
both analytical or numerical statistical mechanics techniques. These
processes are of fundamental interest since it presents detailed aspects of
the electrode-electrolyte interface structure. Electrical current and
potential for electrochemical deposition in metallic surfaces are obtained
by cyclic voltammetry, showing current peaks in a determined potential
range. The shape, position and number of these peaks depends on the surface
where the deposition occurs, as well as on the electrolyte nature. The peaks
are related to phase transitions of the adsorbed layers (\cite{Dash},\cite
{Nicholson}), and this aspect makes this electrochemical subject be well
adapted to statistical mechanics studies.

The dynamics of the adsorption of atoms and molecules on metallic surfaces
in electrolytic solutions due to voltage differences - the electrochemical
deposition - can be described, in an effective way, by the lattice-gas
model. In this model, the adsorption of an element (adsorbed specie) changes
the energy of the system due the bound energy of the element to the surface,
and due lateral interactions between adsorbed species. The system is
described by a Hamiltonian that reproduces interaction between nearest
neighbors (but not only first nearest neighbors) and terms relative to an
external field (adsorbed specie electrochemical's potential). The
Hamiltonian is usually treated by using computational simulation, such as
Monte Carlo simulations, in both equilibrium (\cite{Landau}-\cite{Rikvold91}%
) and quasi-equilibrium situations(\cite{Gamboa}-\cite{Rikvold98}),
reproducing surface coverage and voltammetric current profiles in good
agreement with experimental data. Also analytical treatment is employed in
these systems (\cite{Huckaby90}-\cite{Blum96}). However, in contrast with
the first ones, these treatments usually employ some empiric approximations.

The analytical treatment of this model is quite difficult, since it involves
up to fifth neighbors interactions \cite{Rikvold98}, characterizing the
non-local nature of the model. Another aspect that increases the complexity
of these treatments arises when we consider potential terms of order higher
than quadratic. In fact, those higher order potential terms are usually
avoided.

The continuous limit of the discrete lattice-gas Hamiltonian is a field
theoretical model. In the treatment of such model, there's no great
difficulties, although requiring perturbative expansions, in dealing with
higher order potential terms. Nevertheless, non-local terms in field
theoretical models are of difficult treatment, and as long as it can be
done, one looks always for local field theoretical models.

In this work we present a field theoretical model constructed from the
continuous version of the electrochemical deposition lattice-gas
Hamiltonian. To avoid non-local terms, we will consider only first nearest
neighbors interactions, but, in order to reproduce the phase transition
presented by the electrochemical deposition, we will introduce a fourth
order potential term, leading to the $\lambda \phi ^4$ model.

We must remark, nevertheless, as indicated in \cite{Brown98}, that the
numerical Monte Carlo treatment of such systems has some advantages in the
detailed description of the surfaces, of the distribution and sizes of
deposited ions and molecules, and so on. But an analytical, or specifically,
a field theoretical description of the electrochemical deposition process
can improve the understanding of the phenomena involved, since it provides
simple analytical expressions for some of the observable quantities and
gives a comprehensible description of how the symmetry breaking is related
to the phase transition of the system. It also opens the possibilities for
the study of the presence of topological solutions (solitons and vortices)
in the distribution of the adsorbed specie over the electrode surface,
which, in future, could represent the possibility of new technological
features.

\section{Continuous Version of the Lattice-Gas Model}

Let us start from the grand-canonical lattice-gas Hamiltonian (\cite
{Rikvold88},\cite{Collins},\cite{Rikvold98}) for the electrochemical
deposition of one specie in a square lattice, given by 
\begin{equation}
H=-J\sum_{i,j=1}^N\{c_{i,j}c_{i,j+1}+c_{i,j}c_{i+1,j}\}+\mu
\sum_{i,j=1}^Nc_{i,j}+H_3,  \label{LatticeHamiltonian}
\end{equation}
where J is the lateral interaction between species occupying nearest
neighbors sites, $c_{i,j}=\{0,1\}$ denotes the occupation of sites labeled
by its row and column numbers $i$ and $j$, $\mu $ is the electrochemical
potential and $H_3$ is higher order interaction terms. The electrochemical
potential $\mu $ is related to the electrode potential through: 
\begin{equation}
\mu =\mu _0+kT\ln \frac C{C_0}-e\gamma V,  \label{ecpotential}
\end{equation}
where $C_0$ is some reference value for the specie concentration in
solution, $\mu _0$ is the electrochemical potential at the reference
concentration and zero electrode potential $V$, $C$ is the specie
concentration in solution and $\gamma $ is the electrovalence of the specie.

The thermodynamic conjugate to the electrochemical potential is the surface
coverage $\Theta $, that represents the fraction of occupied sites 
\begin{equation}
\Theta =N^{-1}\sum_{i=0}^Nc_i  \label{surfacecoverage}
\end{equation}
, where $N$ is the total number of unity cells in the electrode surface. The
charge transported through the substrate/electrode interface in
adsorption/desorption process is given by 
\begin{equation}
q=-e\gamma \Theta ,  \label{charge}
\end{equation}
and the voltammetric current density (voltammetric current per adsorption
site) can be obtained from its derivative 
\begin{equation}
i=\frac{dq}{dt}=-e\gamma \frac{d\Theta }{dt}=-e\gamma \frac{d\theta }{d\mu }%
\frac{d\mu }{dV}\frac{dV}{dt},  \label{current}
\end{equation}
where we express eq. (\ref{current}) as a function of the scan rate $\frac{dV%
}{dt}$.

Rewriting (\ref{LatticeHamiltonian}) by using a symmetric occupation number $%
\phi _{i,,j}$, given by 
\begin{equation}
\phi _{i,j}=2c_{i,j}-1;\,\,\,\,\phi _{i,j}=\{-1,+1\}  \label{d_c}
\end{equation}
we find 
\begin{equation}
H=-\frac J4\sum_{i,j=1}^N(\phi _{i,j}\phi _{i,j+1}+\phi _{i,j}\phi
_{i+1,j})-\left( -J+\frac \mu 2\right) \sum_{i,j=1}^N\phi _{i,j}+H_3
\end{equation}
where we had made use of periodic boundary conditions. Introducing the
interaction strength by unitary cell $\frac J{a^2}$, where $a$ is the
lattice spacing and $a^2$ is the unitary cell area, we find 
\begin{equation}
H=-\frac J{4a^2}\sum_{i,j=1}^Na^2\{\phi _{i,j}\phi _{i,j+1}+\phi _{i,j}\phi
_{i+1,j}\}-\frac{-J+\mu /2}{a^2}\sum_{i,j=1}^Na^2\phi _{i,j}+H_3
\end{equation}
or 
\begin{eqnarray}
H &=&-\frac J{4a^2}\sum_{i,j=1}^Na^2\{(-\frac 12\phi _{i,j}^2+\phi
_{i,j}\phi _{i,j+1}-\frac 12\phi _{i,j+1}^2)+(-\frac 12\phi _{i,j}^2+\phi
_{i,j}\phi _{i+1,j}-\frac 12\phi _{i+1,j}^2\}  \label{hamiltonian2} \\
&&-\frac{Jz}{16a^2}\sum_{i,j=1}^Na^2\{\phi _{i,j}^2+\frac 12\phi _{i,j+1}^2+%
\frac 12\phi _{i+1,j}^2\}-\frac{-J+\mu /2}{a^2}\sum_{i,j=1}^Na^2\phi
_{i,j}+H_3.  \nonumber
\end{eqnarray}
where $z$ is the number of nearest neighbors to the site $i,j$.

Applying the periodic boundary conditions, eq. (\ref{hamiltonian2}) reduces
to 
\begin{eqnarray}
H &=&\frac J8\sum_{i,j=1}^Na^2\left\{ \left( \frac{\phi _{i,j}-\phi _{i,j+1}}%
a\right) ^2+\left( \frac{\phi _{i,j}-\phi _{i+1,j}}a\right) ^2\right\}
\label{hamiltonian3} \\
&&-\frac{Jz}{8a^2}\sum_{i,j=1}^Na^2\phi _{i,j}^2-\frac{-J+\mu /2}{a^2}%
\sum_{i,j=1}^Na^2\phi _{i,j}+H_3.  \nonumber
\end{eqnarray}

In order to obtain a field model that corresponds to (\ref{hamiltonian3}),
we must assume that the lattice spacing is negligible when compared to the
lattice dimensions, and that the occupation number of neighbors sites
changes slowly from one site to another. In fact, as stated by eq. (\ref{d_c}%
), $\phi _{i,j}$ can assume only one of two discrete values, $\pm 1$. To
obtain a genuine field theory, we must replace the occupation number $\phi
_{i,j}$ in eq. (\ref{hamiltonian3}) by its configurational mean $\phi
(x,y)=<\phi _{i,j}>$, where the brackets represents the mean over all
configurations allowed to the system at certain fixed conditions
(temperature, solute concentration, electrode potential). Since cyclic
voltammetry is carried out with slow scan rates, in a quasi-statical
procedure, it seems to be reasonable that all allowed states with some fixed
energy could contribute to a specific current measurement.

Denoting $H$ and $\phi $ for the configurational mean hamiltonian and
occupation number respectively from now on, we can take the limit $%
a\rightarrow 0$, and replace the lattice columns and rows labels $i$ and $j$
by the sites positions $x$ and $y$, resulting in 
\begin{equation}
H=\int dxdy\left\{ \frac J8\left( \overrightarrow{\nabla }\phi \right) ^2-%
\frac{Jz}{8a^2}\phi ^2-\frac{-J+\mu /2}{a^2}\phi \right\} +H_3=\int
dxdy\,\,h,
\end{equation}
where $\phi =\phi (x,y)$. Identifying the Hamiltonian density $h$, we found 
\begin{equation}
h=\frac J8\left( \overrightarrow{\nabla }\phi \right) ^2-\frac{Jz}{8a^2}\phi
^2-\frac{-J+\mu /2}{a^2}\phi +h_3,
\end{equation}
where $h_3$ is the Hamiltonian density of the higher order interaction
terms, and 
\begin{equation}
H_3=\int dxdy\,\,h_3.
\end{equation}

By applying the appropriate Legendre transformation, we obtain the
Lagrangian density for the continuous model 
\begin{equation}
l=-\frac 12\frac J4\left( \overrightarrow{\nabla }\phi \right) ^2-\frac 12%
\left( -\frac{Jz}{4a^2}\right) \phi ^2+\frac{-J+\mu /2}{a^2\rho }\phi -h_3.
\end{equation}
Here, we are also neglecting higher order correlations, i.e., we are
assuming $<\phi ^2>=<\phi >^2$.

Rewriting the Lagrangian density in terms of 
\begin{equation}
\varphi =\frac{\sqrt{J}}2\phi  \label{fieldterm}
\end{equation}
we obtain 
\begin{equation}
l=-\frac 12\left( \overrightarrow{\nabla }\varphi \right) ^2-\frac 12\left( -%
\frac z{a^2}\right) \varphi ^2+\frac{-2J+\mu }{a^2\sqrt{J}}\varphi -h_3.
\end{equation}

Finally, defining the quantities $m$ and $c$ by 
\begin{equation}
m^2=-\frac z{a^2},  \label{massterm}
\end{equation}
and 
\begin{equation}
c=\frac{2J-\mu }{a^2\sqrt{J}},  \label{breakterm}
\end{equation}
we obtain, for the Lagrangian density of the continuous extension of the
lattice gas Hamiltonian applied to the electrochemical deposition, 
\begin{equation}
l=-\frac 12\left( \overrightarrow{\nabla }\varphi \right) ^2-\frac 12%
m^2\varphi ^2+c\varphi -h_3,  \label{scalarmodel}
\end{equation}
formally identical to a scalar field model with a symmetry breaking term.
The quantities appearing in (\ref{scalarmodel}) are related to the lattice
gas model effective quantities through eq. (\ref{d_c}),(\ref{fieldterm}),(%
\ref{massterm}) and (\ref{breakterm}).

The surface coverage, given by eq. (\ref{surfacecoverage}), can be given in
this continuous version of the lattice-gas model by 
\begin{equation}
\Theta =N^{-1}\sum_{i=1}^Nc_i=N^{-1}J^{-1/2}\sum \varphi _i+\frac 12%
=N^{-1}a^{-2}J^{-1/2}\int dx\,dy\,\varphi (x,y)+\frac 12,
\label{sccontinuous}
\end{equation}
where we have made use of (\ref{d_c}) and (\ref{fieldterm}).

The model described by (\ref{scalarmodel}) corresponds to the non-linear
sigma model, since it is restricted, by means of eq. (\ref{d_c}), to $-1\leq
<\phi >\leq 1$, wich implies in 
\begin{equation}
-\frac{\sqrt{J}}2\leq \varphi \leq \frac{\sqrt{J}}2.
\end{equation}

In this work we are interested in the description of the phase transition in
electrochemical deposition, which occurs in the region far from saturation ($%
\varphi =\pm \sqrt{J}/2$). Thus, we will treat the $\varphi $ as a
non-restricted field, keeping in mind that the results will be valid only in
the region where $|\varphi |<\sqrt{J}/2$.

\section{The $\lambda \varphi ^4$ model in the Mean Field Approximation}

From eq. (\ref{scalarmodel}) we can see that, with vanishing breaking term
and higher order interactions, the Lagrangian for the continuous version of
the lattice gas model presents a symmetry related to the exchange of the
occupation number $\varphi \rightarrow -\varphi $. Since the transition
between the state where the adsorption of the specie is favorated and the
state where desorption is favoured is dictated by the electrode potential,
present in the explicit symmetry breaking term (\ref{breakterm}), we assume
that the higher order interaction terms will be also symmetric to the
exchange of the occupation number, and thus only even order terms must be
present in $h_3$. The lowest even order interaction term is the $\varphi ^4$%
, and we will assume 
\begin{equation}
h_3=\frac \lambda {4!a^2}\varphi ^4\text{,}  \label{h3}
\end{equation}
where $\lambda $ is the coupling constant and $a^2$ was inserted here for
dimensionality reasons. By this relationship, we are assuming that the
potential energy of the system is non-quadratic in $\varphi $. From (\ref
{scalarmodel}) and (\ref{h3}) we obtain 
\begin{equation}
l=-\frac 12\left( \overrightarrow{\nabla }\varphi \right) ^2-\frac 12%
m^2\varphi ^2+c\varphi -\frac \lambda {4!a^2}\varphi ^4,  \label{phi4model}
\end{equation}
the $\lambda \varphi ^4$ model with an explicit symmetry breaking term. Due
the fact that the mass term in (\ref{phi4model}) is positive, since $m^2$ is
negative (\ref{massterm}), the model presents spontaneous symmetry break in
the limit $c=0$. In this limit, the minimum of the potential energy in (\ref
{phi4model}) occurs in 
\begin{equation}
\varphi _{\min }=\sqrt{-\frac{3!m^2a^2}\lambda }\equiv \eta
\end{equation}
or 
\[
\varphi _{\min }=-\sqrt{-\frac{3!m^2a^2}\lambda }. 
\]

Perturbative solutions must be constructed around one of these minima.
Introducing the shifted field $\varphi _s=\varphi -\eta $ in the Lagrangian (%
\ref{phi4model}), we find 
\begin{eqnarray}
l &=&-\frac 12\left( \overrightarrow{\nabla }\varphi _s\right) ^2-\left( 
\frac 12m^2+\frac{6\eta ^2\lambda }{4!a^2}\right) \varphi _s^2
\label{shiftedlagrangian} \\
&&+\left( c-m^2\eta -\frac{\eta ^3\lambda }{3!a^2}\right) \varphi _s-\frac{%
\eta \lambda }{3!}\varphi _s^3-\frac \lambda {4!a^2}\varphi _s^4.  \nonumber
\end{eqnarray}

Lagrangian (\ref{shiftedlagrangian}) is no longer symmetric in the sense
that it is not invariant under the exchange $\varphi _s=-\varphi _s$, so the
model shows an spontaneous symmetry breakdown. The potential energy minimum
occurs now at the point $\varphi _s=0$. Applying the Euller-Lagrange
equations 
\begin{equation}
\vec{\nabla}\left( \frac{\partial l}{\partial (\vec{\nabla}\varphi _s)}%
\right) -\frac{\partial l}{\partial \varphi _s}=0
\end{equation}
in (\ref{shiftedlagrangian}), we obtain, for the dynamic equation of the
system, 
\begin{eqnarray}
-\nabla ^2\varphi _s &=&-\frac \lambda {6a^2}\varphi _s^3-\frac{\lambda \eta 
}{2a^2}\varphi _s^2-2\left( \frac{\lambda \eta ^2}{4a^2}+\frac{m^2}2\right)
\varphi _s  \label{EullerLagrange} \\
&&+\left( -\frac{\eta ^3\lambda }{6a^2}-m^2\eta +c\right) .  \nonumber
\end{eqnarray}

We can proceed the solution of eq. (\ref{EullerLagrange}) perturbativelly
around $\varphi _s=0$. Thus, expanding $\varphi _s$ as a power series of a
perturbative parameter $\varepsilon $, i.e., $\varphi _s=\varphi
_0+\varepsilon \varphi _1+\varepsilon ^2\varphi _2+...$, with $\varphi _0=0$%
, and keeping only the lowest order terms ($\varepsilon ^0$) implies in 
\begin{equation}
-\frac{\eta ^3\lambda }{6a^2}-m^2\eta +c=0  \label{cubic}
\end{equation}
, a cubic equation that can be solved analytically for $\eta $. As showed in
figure 1, equation (\ref{cubic}) presents one (figures 1a I and III) or
three real solutions (figure 1a II), for the potential energy as a function
of $\varphi _s$, corresponding to the local minimum or maximum. Stability is
granted for the global minimum, for $c=0$ we have two symmetric global
minima, but for $c\neq 0$ the system has this symmetry explicitly broken, as
one can see from figures 1b - 1d. Thus, in $c=0$ the system presents a
transition between one global minimum to another.

Solutions to equation (\ref{cubic}) is given analytically by 
\begin{equation}
\eta _1=\frac{\Delta ^{1/3}}\lambda -\frac{2m^2a^2}{\Delta ^{1/3}},
\label{eta1}
\end{equation}
\begin{equation}
\eta _2=\left( -\frac{\Delta ^{1/3}}{2\lambda }+\frac{m^2a^2}{\Delta ^{1/3}}%
\right) +\sqrt{3}i\left( \frac{\Delta ^{1/3}}{2\lambda }+\frac{m^2a^2}{%
\Delta ^{1/3}}\right)  \label{eta2}
\end{equation}
and 
\begin{equation}
\eta _3=\left( -\frac{\Delta ^{1/3}}{2\lambda }+\frac{m^2a^2}{\Delta ^{1/3}}%
\right) -\sqrt{3}i\left( \frac{\Delta ^{1/3}}{2\lambda }+\frac{m^2a^2}{%
\Delta ^{1/3}}\right) ,  \label{eta3}
\end{equation}
with 
\begin{equation}
\Delta =3ca^2\lambda ^2+\lambda ^2\sqrt{\frac{8m^6a^6+9c^2a^4\lambda }\lambda
}.
\end{equation}

We must observe that solutions (\ref{eta1}), (\ref{eta2}) and (\ref{eta3})
are independent of the unitary cell area, since it appears only in the
combinations $m^2a^2$ and $ca^2$, and, due (\ref{massterm}) and (\ref
{breakterm}) 
\begin{equation}
m^2a^2=-\frac z{a^2}a^2=-z
\end{equation}
and 
\begin{equation}
ca^2=\frac{2J-\mu }{a^2\sqrt{J}}a^2=\frac{2J-\mu }{\sqrt{J}}.
\end{equation}
Thus, the number of free parameters of the model is reduced for only four:
the electrochemical potential $\mu $, the effective lateral interaction $J$,
the strength of the non-quadratic local interaction $\lambda $ and the
effective electrovalence of the adsorbed specie $\gamma $. The latter is
also limited into a range of values near the electrovalence of the adsorbed
specie.

First order ($\varepsilon ^1$) solutions to (\ref{EullerLagrange}) implies
in 
\begin{equation}
\nabla ^2\varphi _1=\left( \frac{\lambda \eta ^2}2+m^2\right) \varphi _1.
\label{phi1}
\end{equation}

Solutions of eq. (\ref{phi1}) can improve the results presented in this
work, with oscilatory solutions around the mean solution $\varphi _0$.

\section{Soft Break Limit}

In the limit of vanishing $c$, solutions to eq. (\ref{cubic}) reduces to 
\begin{equation}
\eta _{c=0}=0;\,\,\eta _{c=0}=\pm \sqrt{\frac{-6m^2a^2}\lambda }.
\label{etac0}
\end{equation}

For very small $c$, we must suppose $\eta =\eta _{c=0}+\delta $, where $%
\delta $ is a small perturbation around the $c=0$ solution. Replacing these
perturbative solution in eq.(\ref{cubic}) and retaining only first order
terms in $\delta $, one can find 
\begin{equation}
\eta _{\pm }=\pm \sqrt{\frac{-6m^2a^2}\lambda }-\frac c{2m^2}.
\label{etapert}
\end{equation}

This approximate solution for $\eta $ allows us to obtain some simple
estimative expressions for the voltammetric current and surface coverage
that guide us in the initial parameters choice for the non-linear fitting of
the model to the experimental data. The physical measurable parameters are
the electrode potential $V$, the scan rate $dV/dt$, the voltammetric current 
$i$ and the surface coverage $\Theta $.

The energy of the system in the state represented by solution (\ref{etapert}%
) can be read from eq. (\ref{shiftedlagrangian}). The solution that
represents the global minimum energy switches from $\eta _{+}$ to $\eta _{-}$
when $c$ runs from negative to positive values. The electrode potential at
the transition point, obtained from (\ref{breakterm}) and (\ref{ecpotential}%
) for $c=0$ is given by 
\begin{equation}
V_c=\frac{-2J+\mu _{0C}}{e\gamma },  \label{criticalv}
\end{equation}
where we call $\mu _{0C}=\mu _0+kT\ln (C/C_0)$, the electrochemical
potential at solute concentration $C$ and temperature $T$. The current at
the transition point can be obtained from (\ref{current}), where the
intermediate derivatives can be obtained from (\ref{breakterm}) and (\ref
{ecpotential}), and, from (\ref{cubic}), we have 
\begin{equation}
-\frac \lambda {2a^2}\eta ^2d\eta -m^2d\eta +dc=0
\end{equation}
i.e. 
\begin{equation}
\frac{d\eta }{dc}=\frac 2{\lambda m^2/a^2+2m^2}.  \label{detadc}
\end{equation}

Replacing these results in (\ref{current}), we obtain 
\begin{equation}
i=\frac{e^2\gamma ^2}J\frac 2{\lambda \eta ^2+2m^2a^2}\frac{dV}{dt},
\label{current2}
\end{equation}
a result that is not restricted to the soft break limit.

At the transition point, where $\eta $ is given by the non zero solutions of
(\ref{etac0}), replacing (\ref{massterm}), we obtain 
\begin{equation}
i_C=\frac{e^2\gamma ^2}{8J}\frac{dV}{dt}.  \label{criticalcurrent}
\end{equation}

Near $c=0$, with $\eta $ given by (\ref{etac0}), $i^{-1}$ is quadractic in $%
V $, as one can see from (\ref{current2}), (\ref{etapert}),(\ref{breakterm})
and (\ref{ecpotential}). The critical electrode potential can be evaluated
by means of eq. (\ref{ecpotential}) and (\ref{breakterm}) with $c=0$.

The fact that $i^{-1}$ is quadractic in $V$ near the critical point,
together with eq. (\ref{criticalv}) and (\ref{criticalcurrent}) can be used
to set the initial guess parameters for a non-linear fitting of eq. (\ref
{current2}) to experimental data.

\section{Numerical Results}

Voltammetric current and surface coverage for the process of
electrodeposition of one single adsorbed specie on a square-lattice metallic
surface at constant temperature and solute concentration can be reproduced
in lowest order of the mean field approximation by equations (\ref{current2}%
) and (\ref{sccontinuous}) . The free parameters of the model are $J$, $\mu
_{0C}$, $\lambda $ and $\gamma $. The latter one is restricted to values
near the electrovalence of the adsorbed specie. In comparison with the usual
Monte Carlo approach, it is a reduced set of parameters, since in this
approach, besides the parameters $\gamma $ and $\mu _{0C}$, up to five
lateral interaction between species is used in general \cite{Rikvold98}.

Figure 2 shows the results of the fitting of eq. (\ref{current2}), with $%
\eta $ given by (\ref{eta1}) to (\ref{eta3}), to experimental and simulated
data for the electrosorption of $Br$ in the surface of $Ag(100)$ electrodes (%
\cite{Mitchell},\cite{Ocko},\cite{Valette}). We use a non-linear least
squares fitting, applying the steepest descent procedure. Initial values for
the parameters are estimated using eq. (\ref{criticalv}) and (\ref
{criticalcurrent}), and the $Br$ electrovalence. The best fit results in $%
\lambda =-47.7\times 10^4$, $\mu _{0C}=610meV$ and $J=44.6meV$. Allthough $J$
and $\mu _{0C}$ are strongly dependent on $\gamma $, the set of parameters
gives a fit of the ratio $i/(dV/dt)$ that is independent of $\gamma $. The
surface coverage, neverthelles, depends on the value of the effective
electovalence, and it was used to fix $\gamma =-0.73$, which corresponds to
the value given in \cite{Mitchell} and gives a good agreement between the
model and experimental data\cite{Ocko}. This value is also near the
effective electrovalence estimated as $\gamma =-0.71$ in \cite{Hamad}, and
no significative differences are found in the results computed with $\gamma
=-0.73$ or $\gamma =-0.71$. Experimental CVs present a shoulder followed by
a sharp peak, this second corresponds to a phase transition from a
disordered to an ordered phase \cite{Ocko}. The shoulder is atributed to
configurational flutuations on the $Br$ adsorbed layer \cite{Mitchell}, and
of course cannot be reproduced in the $\phi ^4$ model. Due the size of
adsorbed bromine being bigger than the $Ag(100)$ lattice spacing, in the
ordered phase the bromine is distributed in an arrangement that covers only
50\% of the $Ag$ surface. Results for the surface coverage presented in
figure 3 are normalized to reproduce this fact, and shows an reasonable
agreement with the experimental data.

Figure 4 shows the results of the fitting of the model to experimental data
for the Urea over Pt(100) surface \cite{Rikvold96}. Fitted parameters are $%
\lambda =-58.4\times 10^4$, $\mu _{0C}=97.8\,meV$, $\gamma =-1$ and $%
J=3.04meV$ . Figure 5 shows the surface coverage for this process. In figure
4 the dotted line corresponds to the quadratic fitting of $i^{-1}$. We can
see that there is an exceptional agreement between experimental data and the
quadratic fitting in the sub critical potential region. Of course, this
result is beyond the model presented here, since this quadratic solution do
no take into account higher order terms in $c$ on the solution of eq. (\ref
{cubic}).

On figure 5 the dashed horizontal lines correspond to the limits $<\varphi
>=\eta =\pm \sqrt{J}/2$, values of $\eta $ inside this region and relatively
far from these limits corresponds to the linear approximation of (\ref
{scalarmodel}). As one can see from figures (4) and (5), the phase
transition region is inside these limits. In particular, for the $Br$
adsorption in $Ag$, all the voltammogram depicted on figure 2 is inside the
region that corresponds to the linear approach.

Both results shows a good agreement between calculated and
experimental/simulated data. The computational advantage of the field
theoretical approach over the Monte Carlo simulation is the reduced number
of parameters of the former.

\section{Conclusion}

In conclusion, we constructed a field theoretical model to describe
electrochemical deposition. Based on the lattice-gas hamiltonian of the
process, and taking a configurational mean of the occupation numbers, we
obtained, in the continuous limit, a scalar field hamiltonian whose
coefficients are related to the lattice-gas model effective quantities.
Including higher order potential and treating the model in the mean field
approximation, we obtain a single system of equations ((\ref{sccontinuous}),
(\ref{eta1})-(\ref{eta3}) and (\ref{current2})) that reproduces the current
versus electrode potential behaviour in electrochemical deposition. The
fitting of the model with experimental and numeric simulated data was
obtained and shows a good agreement between model and experimental/simulated
results. In comparison with the Monte Carlo approach, the model presented
here has the advantage of presenting a reduced set of free parameters and
simple analytical expressions for the observed quantities.

These results open the possibilities of new analytical description of other
more complex deposition process, such as two species deposition, where one
must use a two scalar field model, introducing new symmetries in the model
(work in progress). Other even order potential terms could also be used to
reproduce phase transitions between different ordered phases. They also
suggests that the non-linear description of the process can presents other
solutions, such as topological solutions like solitons and vortices. In
fact, the $\lambda \phi ^4$ term can be viewed as a fourth order term in the
expansion of a $\cos (\phi )$ term in the lagrangian (\ref{scalarmodel}),
leading to the sine-gordon model \cite{sinegordon}, which presents solitons.

Finally, the hamiltonian studied in this work shows us that the phase
transition presented by electrochemical deposition is related to an
explicitly symmetry breaking process, dictated by an external field
(electrochemical potential), controlled by the electrode potential.

\section{Acknowledgments}

The authors are grateful to CAPES-Brazil. We also thanks A. R. Pereira and
M. C. Nemes for their suggestions to the present work.

%
%

\newpage

\begin{quote}
{\bf Figure Captions\\}
\end{quote}

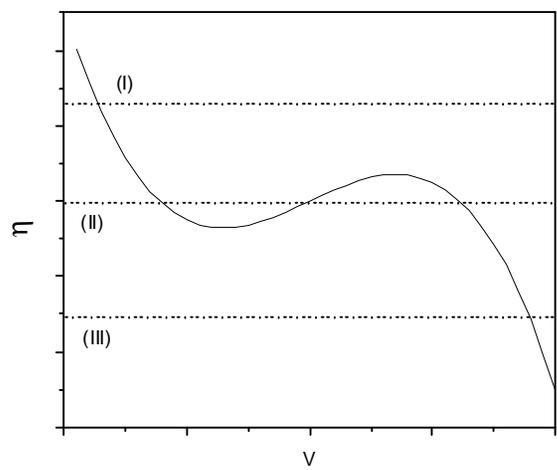
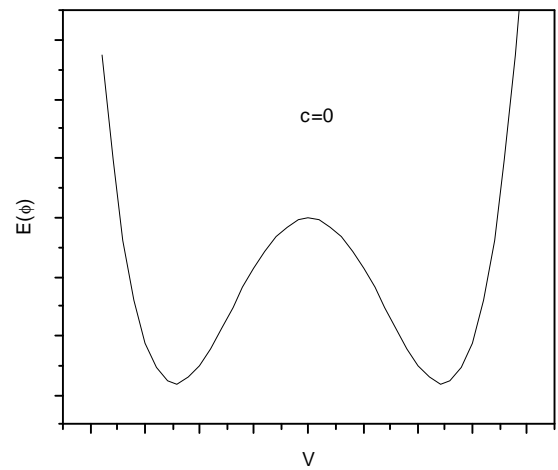
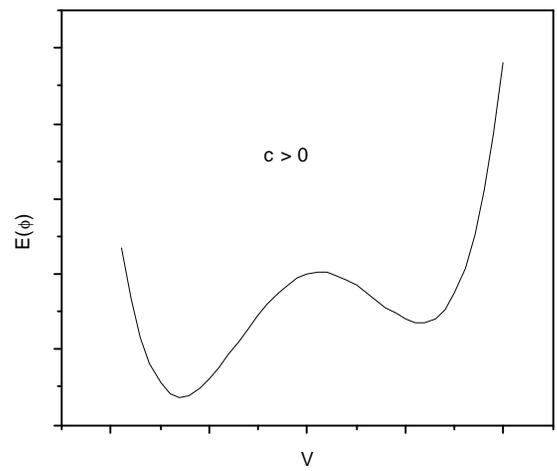
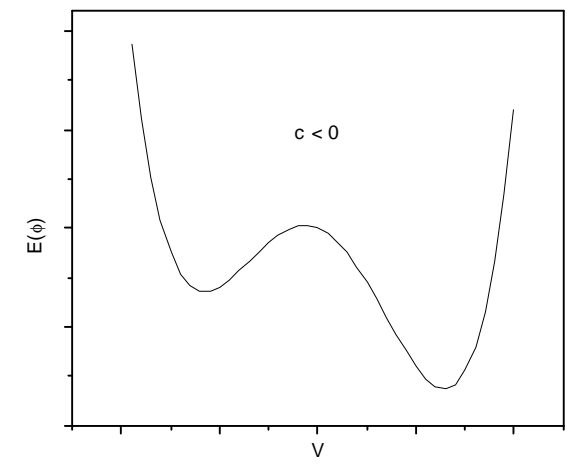
Figure 1: (a) Cubic equation corresponding for the configurational mean
value of the occupation number $\phi $. The horizontal dashed lines
correspond to the $\eta =0$ axis with I) c%
\mbox{$<$}
0, II) c=0 and III) c%
\mbox{$>$}
0. Note that for c%
\mbox{$>$}
0 and c%
\mbox{$<$}
0 one can have one or three real solutions. (b), (c) and (d) Potential
energy as a function of $\eta $ for different values of the symmetry
breaking term. For c=0 the two minima are symmetric, as c goes from negative
to positive values, the global minimum switches from one side to another,
corresponding to the phase transition of the system.\\

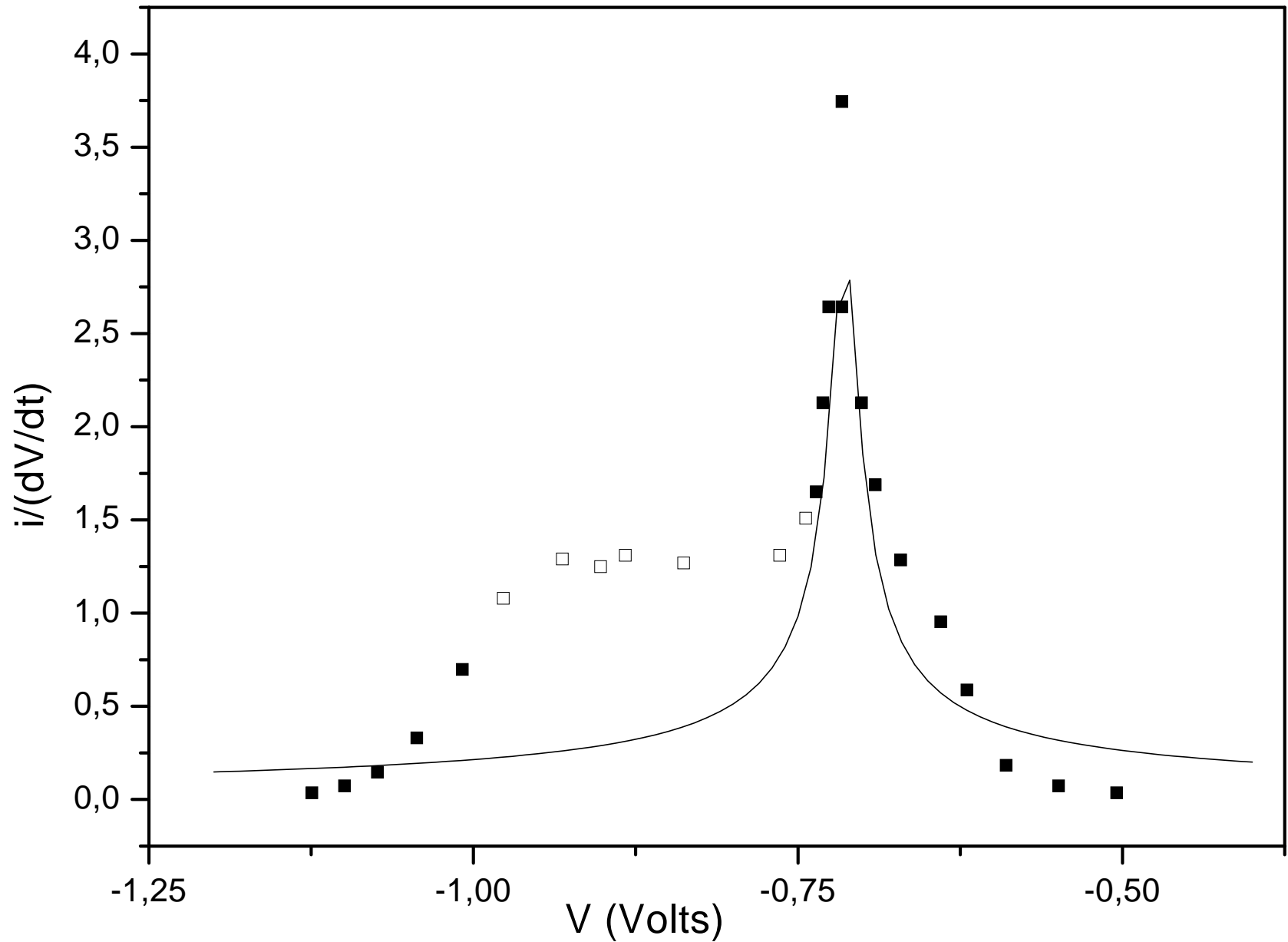
Figure2: Current density as a function of the electrode potential for the
electrodeposition of $Br$ on $Ag(100)$ in the $\phi ^4$ model (solid line).
Full and empty squares correspond to simulated results \cite{Mitchell},
empty squares are points not took into account in the nonlinear fitting.\\

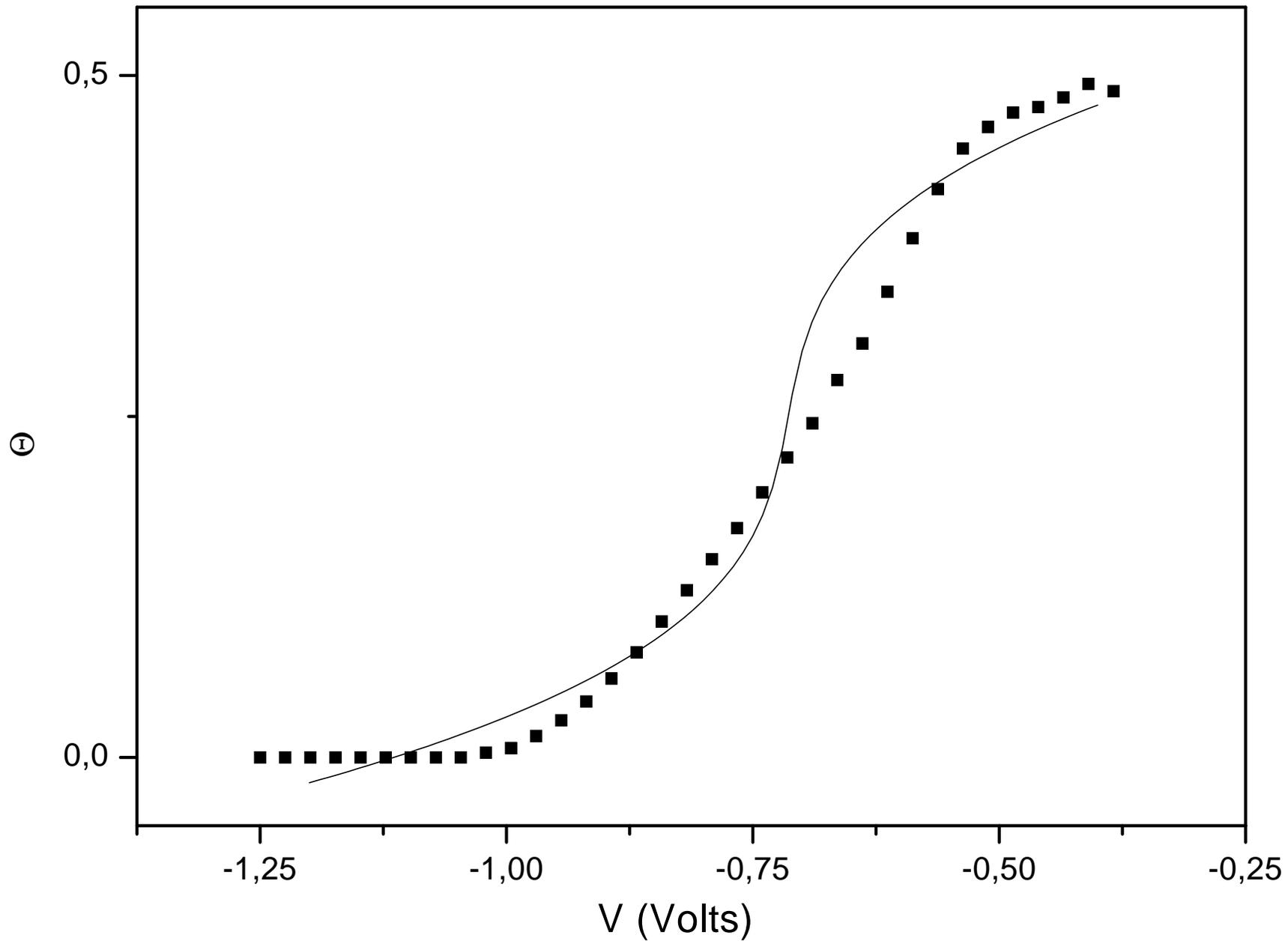
Figure 3: $\phi ^4$ model results (solid line) and experimental results
(full squares) \cite{Ocko} for the surface coverage in the electrodeposition
of $Br$ on $Ag(100)$.\\

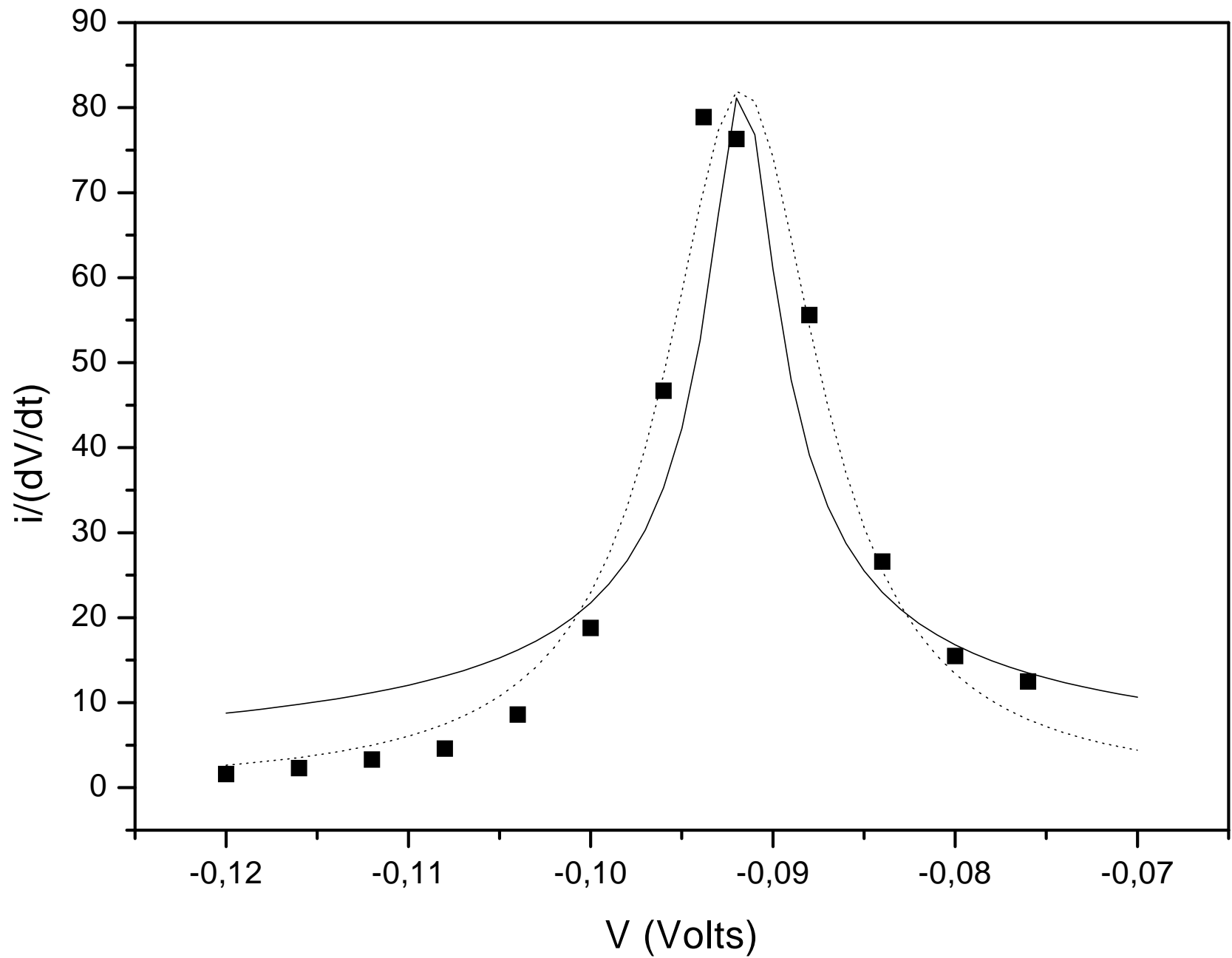
Figure 4: Current density as a function of the electrode potential for the
electrodeposition of Urea on $Pt(100)$ in the $\phi ^4$ model (solid line).
Experimental results (full squares) are obtained from \cite{Rikvold96}.
Dotted line corresponds to a quadratic polynomial fitting of $i^{-1}$.\\

Figure 5: Surface coverage for the the electrodeposition of Urea on $Pt(100)$%
. Dashed horizontal lines correspond to the approximate limits of the linear
approach presented in this work.

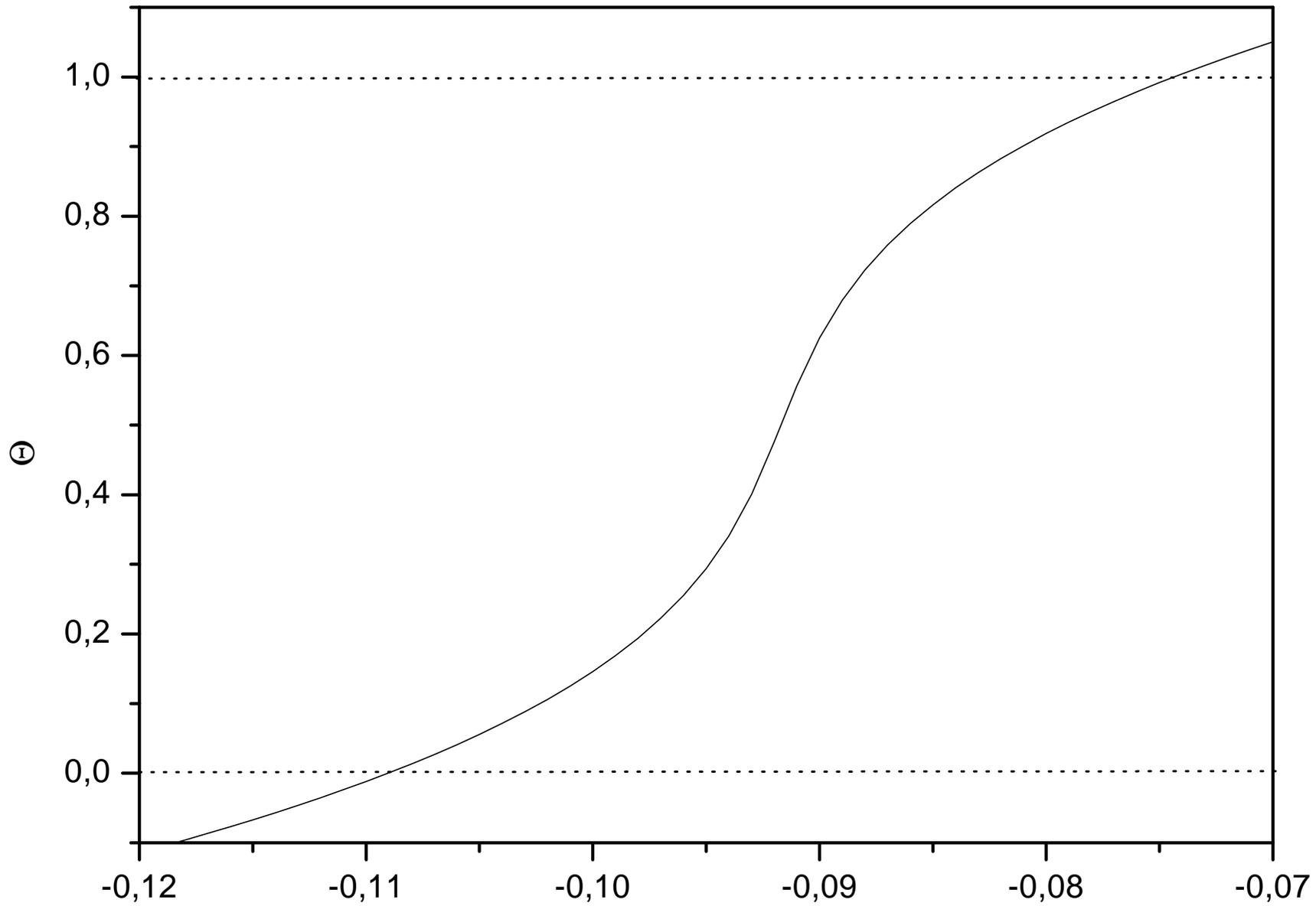
\end{document}